%% file: prl.tex
\documentclass[aps,prl,twocolumn,showpacs,groupedaddress]{revtex4}
\usepackage{graphicx}  
\usepackage{dcolumn}   
\usepackage{bm}        
\usepackage{amssymb}   

\usepackage[usenames]{color}

\input epsf
\newcommand{\MET}{\mbox{$\not\hspace{-0.11cm}E_T$}}

\begin{document}

\hspace{5.2in} \mbox{FERMILAB-PUB-07/499-E}

\title{Search for $W'$ bosons decaying to an electron and a neutrino with the D0 detector}
\input list_of_authors_r2.tex  
\date{October 16, 2007}

\begin{abstract}
This Letter describes the search for a new heavy charged gauge boson $W'$ decaying 
into an electron and a neutrino. The data were collected with the D0 detector at 
the Fermilab Tevatron $p\bar p$ Collider at $\sqrt{s}= 1.96$~TeV, 
and correspond to an integrated luminosity of about 1~fb$^{-1}$. Lacking any 
significant excess in the data in comparison with known processes, an upper limit 
is set on $\sigma _{W'} \times B(W'\rightarrow e\nu)$, and a $W'$ boson with mass below 
1.00~TeV can be excluded at the 95\% C.L., assuming standard-model-like 
couplings to fermions. This result significantly improves upon previous limits, 
and is the most stringent to date.
\end{abstract}

\maketitle

The standard model (SM) describes the fundamental fermions and their interactions 
via gauge bosons at a high level of accuracy, but it is not considered to be a 
complete theory. Additional gauge bosons are introduced in e.\,g. 
left-right-symmetric models (broken $SU(2)_{L} \times SU(2)_{R}$) or in grand 
unified theories which may also involve supersymmetry (e.\,g. $E_6$) \cite{Moha}.
Assuming the most general case, a new gauge group can comprise a new mixing angle 
$\xi$, new couplings to the fermions $g'$ and a new CKM matrix $U'$. In some models 
the $W'$ boson ($W'^+$ or $W'^-$) is right-handed, and decays therefore into a 
right-handed neutrino and a charged lepton. However, such a neutrino has not yet 
been observed. The mass limits on the $W'$ boson will in general depend on 
$\xi$, $g'$, $U'$, and the masses of possible additional neutrinos. 

In this Letter we make the assumption that there is no mixing, $g'$ is equal to 
the SM coupling, $U'$ is equal to the SM CKM matrix, and that the decay channel 
$W'\rightarrow WZ$ is suppressed. Furthermore, the width $\Gamma _{W'}$ of the 
$W'$ boson is assumed to scale with its mass $m_{W'}$, 
\begin{equation}
\Gamma_{W'} = \frac{4}{3}\cdot\frac{m_{W'}}{m_W}\cdot \Gamma_W.
\end{equation}
The factor of 4/3 is applied in order to account for the decay into the third 
quark family (e.\,g. $W'\rightarrow t\bar b$) which is possible for $m_{W'}$ 
above the kinematic threshold for this process. In case of the existence of 
additional generations of fermions, it is assumed that they are too heavy to 
be produced by a $W'$ decay. This generic model has been introduced by 
Altarelli et al. \cite{Altarelli}. It corresponds to the manifest left-right 
symmetric model \cite{LR} with light right-handed neutrinos if the $W'$ boson 
is right-handed. In this report, the general approach \cite{Altarelli} is 
considered, where the additional gauge boson $W'$ can be right- or left-handed. 

The $W'$ boson has been searched for previously by the D0 \cite{d0WR, oldWp, d0Wtb} 
and the CDF experiments \cite{cdfwp1, cdfwp3, cdfwp2} in various final states. 
The most restrictive limit so far is $m_{W'} > 800$~GeV at the 95\% C.L. \cite{oldWp} 
reported by D0 ($W'\rightarrow q\bar q'$, Run~I). 

Data collected with the D0 detector \cite{d0det} at the Fermilab Tevatron $p\bar p$ 
Collider at a center-of-mass energy of 1.96~TeV are analyzed for the production 
of $W'$ bosons and the subsequent decay into an electron and a neutrino. The neutrino 
can not be detected, but it gives rise to missing transverse energy ($\MET$) in the 
detector. The dataset corresponds to an integrated luminosity \cite{d0lumi} of 
0.99~$\pm$~0.06~fb$^{-1}$, and was collected between 2002 and 2006 during Run~II 
of the Tevatron.

The D0 detector has a central-tracking system, consisting of a 
silicon microstrip tracker (SMT) and a central fiber tracker (CFT), 
both located within a 2~T superconducting solenoidal 
magnet, with designs optimized for tracking and 
vertexing at $|\eta|<3$ and $|\eta|<2.5$, respectively,
where $\eta=-\ln\tan\frac{\theta}{2}$ is the pseudorapidity and $\theta$ 
the polar angle w.r.t. the $z$-axis (proton-beam direction).
A liquid-argon and uranium calorimeter has a 
central section (CC) covering pseudorapidities $|\eta|$ up to 
$\approx 1.1$, and two end calorimeters (EC) that extend coverage 
to $|\eta|\approx 4.2$, each housed in separate 
cryostats. An outer muon system, at $|\eta|<2$, 
consists of a layer of tracking detectors and scintillation trigger 
counters in front of 1.8~T iron toroids, followed by two similar layers 
after the toroids. 
Luminosity is measured using plastic scintillator 
arrays placed in front of the EC cryostats. 

Different SM processes contribute to the electron and $\MET$ final 
state: inclusive production of $W$ or $Z$ bosons, di-bosons ($WW$, $WZ$, $ZZ$)
or $t\bar t$ pairs where at least one boson or one top quark decays
into electrons directly or via tau decays.
In these processes the missing energy is due to the neutrino.
There are also two sources of misidentification background that can contribute 
to the electron and $\MET$ final state: QCD multijet background with one jet 
misidentified as an electron and energy mismeasurement which can 
cause large $\MET$ either along or in the opposite direction of the electron, and
$Z\rightarrow ee$ events where one electron is lost (e.\,g. entering 
non-instrumented sections of the calorimeter) or misreconstructed. 
The latter case can lead to large $\MET$. 

The $W'$ signal and SM processes (including $Z\rightarrow ee$) have been 
simulated with the \textsc{pythia} 6.323 \cite{pythia} Monte Carlo program 
using the CTEQ6L1 \cite{cteq} parton distribution functions (PDFs), except 
for the QCD multijet background, which is estimated from data. 
The generated events are passed through a detailed detector simulation based 
on \textsc{geant} \cite{geant}, and combined with randomly triggered events 
from data to simulate the effects of pile-up and multiple interactions. 
Higher order corrections to the \textsc{pythia} leading order cross sections 
($K$ factors) have been applied. The next-to-next-to-leading order (NNLO) $K$ 
factors and errors due to PDF uncertainties for the signal, the $W$ and the $Z$ 
samples are extracted from Ref. \cite{hamberg}; the NNLO (NLO) cross section 
for $t\bar t$ (di-boson) production is taken from Ref. \cite{ttbar} (\cite{diboson}). 

\begin{figure}
\def\epsfsize#1#2{0.3#1} 
\hspace{\fill}\epsffile{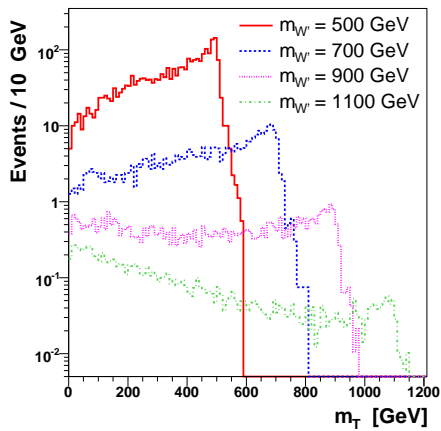} \hspace{\fill}
\caption{Transverse mass $m_T$ distributions for different masses of 
the $W'$ boson (generator level, \textsc{pythia}). The event numbers correspond 
to an integrated luminosity of 1~fb$^{-1}$.\label{fig:signal_mc} }
\end{figure}

The signal cross section falls steeply with increasing mass of the $W'$ boson. 
In addition, for very large masses the on-mass-shell production of $W'$ bosons 
is heavily suppressed due to the smallness of the PDFs at large $x$. 
As shown in Fig. \ref{fig:signal_mc}, the Jacobian distribution no longer 
exhibits a pronounced peak. The transverse mass $m_T$ is calculated from the 
transverse energy of the electron, $E_T^{\rm el}$, the missing 
transverse energy, $\MET$, and the azimuth angle \cite{phi} difference between 
the electron and $\MET$ via
\begin{equation}
m_T = \sqrt{2E_T^{\rm el} \MET(1 - \cos{\Delta \phi(\mbox{electron, $\MET$})})}.
\end{equation}

Events triggered by a set of inclusive single electron triggers are considered.
Electrons with $E_T^{\rm el}>30$~GeV passing the offline identification criteria 
are selected. Monte Carlo studies have shown that the majority ($\approx 80$\%) 
of the electrons stemming from the $W'$ decays are emitted into 
the central detector region. Since the forward detector region exhibits a 
small signal-to-background ratio, only electrons reconstructed in the CC are 
used in the analysis. Electromagnetic clusters are built around a calorimeter 
seed. Such clusters consist of cells in a cone 
($\Delta R = \sqrt{(\Delta \eta)^2 + (\Delta \phi)^2}<0.4$) 
around the seed. Furthermore, the electron shower is required to be isolated in 
the calorimeter, and to deposit most of its energy ($>$~90\%) in the electromagnetic 
part of the calorimeter. The isolation 
$I = [E_{\rm tot}^{0.4}-E_{\rm EM}^{0.2}]/E_{\rm EM}^{0.2}$, which
uses the total shower energy, $E_{\rm tot}^{0.4}$, in a cone of radius $R=0.4$
and the electromagnetic energy, $E_{\rm EM}^{0.2}$, in a cone of radius $R=0.2$, 
is required to be less than 0.2. A cut on the electron shower shape variable is 
applied to separate electromagnetic from hadronic showers. The electron is 
required to have a track matched in $z$ and $\phi$ direction and to stem from 
the primary vertex. Correction factors are applied in order to take differences 
in the reconstruction efficiencies observed in data and Monte Carlo into account.
Finally, the energy dependence of the basic electron reconstruction criteria has 
been studied with simulated electrons from $W'$ decays. The reconstruction 
efficiency is found to be constant (94~$\pm$~1~\%) and does not exhibit a visible 
energy dependence within the statistical uncertainties of the Monte Carlo samples.
The $\MET$ is calculated from all calorimeter cells. Corrections are applied to 
account for the electromagnetic and jet energy scales. We require $\MET>30$~GeV.

Since the transverse momentum of the neutrino is expected to be balanced by the 
electron transverse energy in signal events, a selection on the ratio of the 
energies is applied, $0.6<E_T^{\rm el}/\MET<1.4$. This requirement reduces 
in\-stru\-mental backgrounds from misidentified $\MET$. Jets are reconstructed 
with the iterative mid-point cone algorithm ($R=0.5$) \cite{d0jets}. 
If any jets with $p_T>15$~GeV are present in the event, we require 
$\Delta \phi$(jet, electron)~$<$~2.8 and $\Delta \phi$(jet,~$\MET$)~$<$~2.8. 
These selections remove events from QCD multijet production.

\begin{figure*}
\def\epsfsize#1#2{0.295#1}
\epsffile{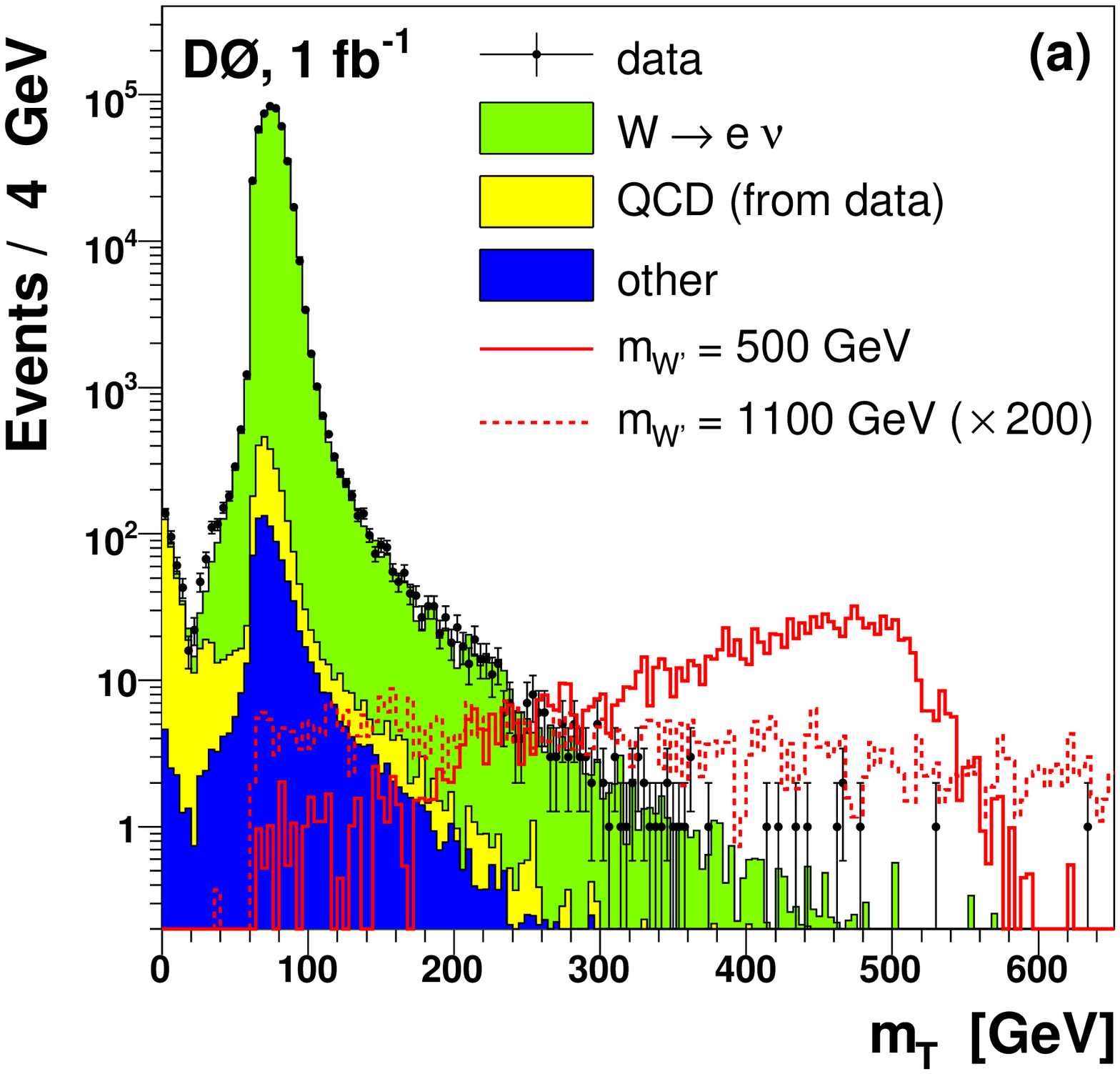} \epsffile{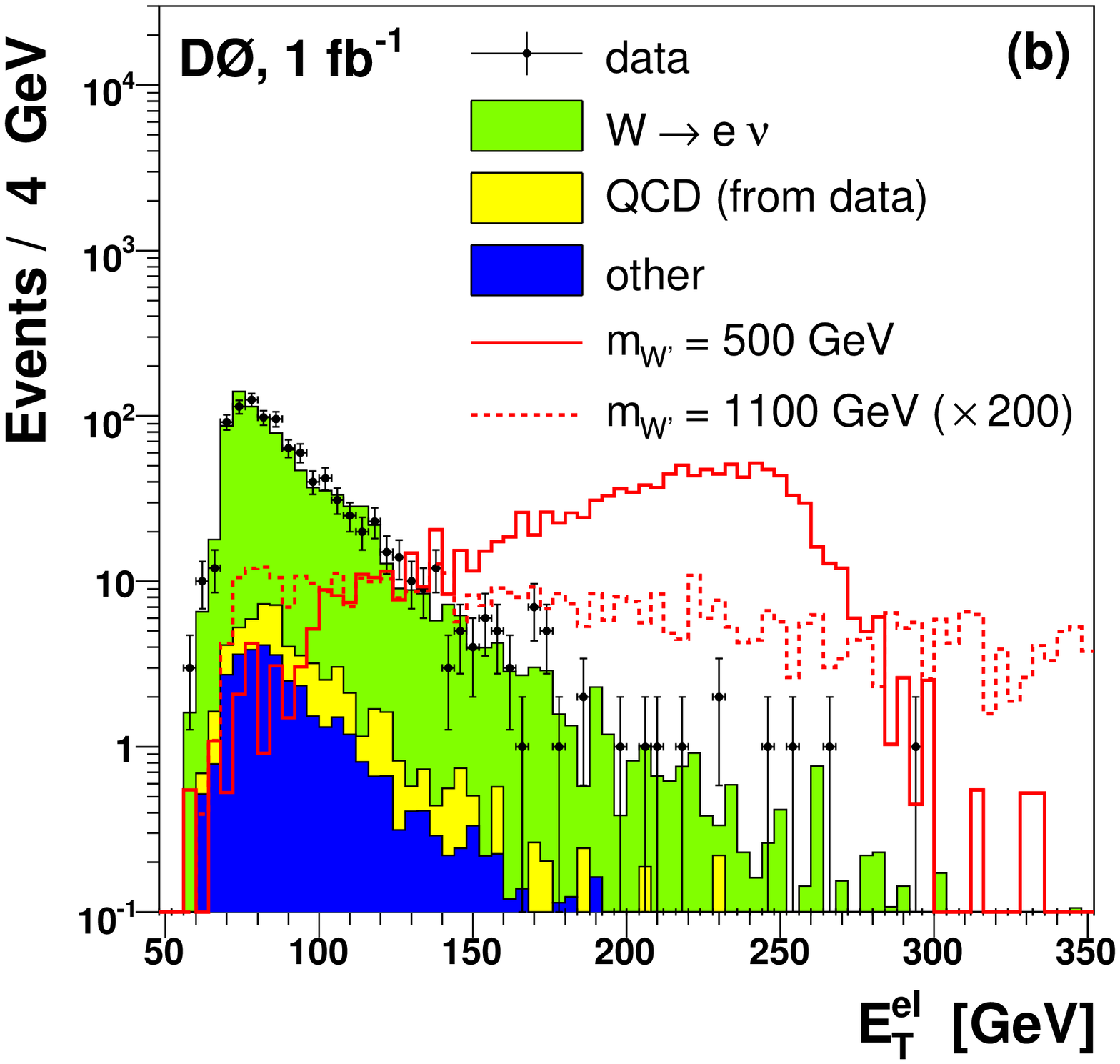} \epsffile{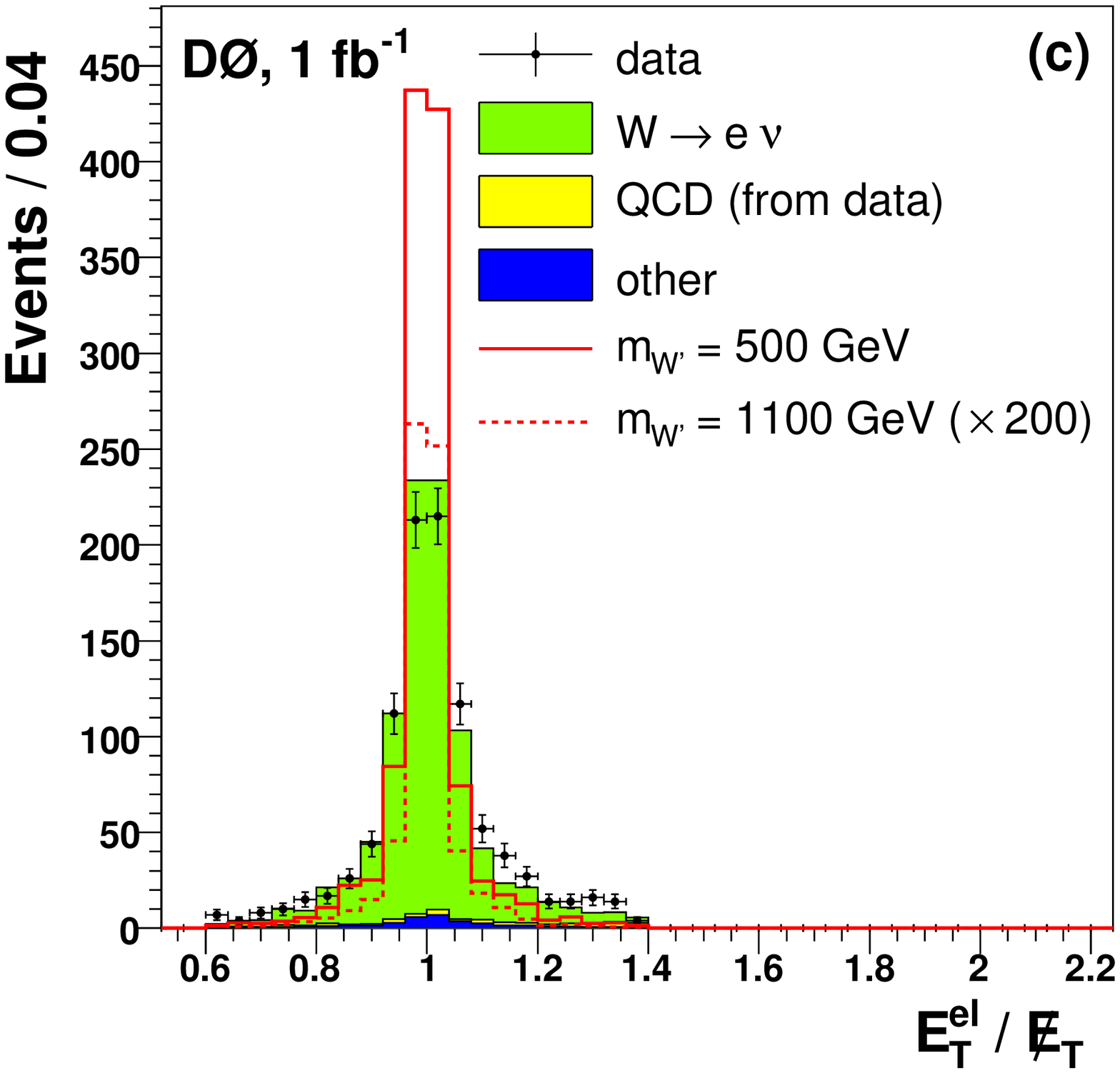}
\caption{Comparison between data and background prediction:
(a) Distribution of the transverse mass $m_T$; (b) distribution of the electron 
transverse energy $E_T^{\rm el}$ in events with $m_T>140$~GeV; 
(c) distribution of the ratio of electron transverse energy and $\MET$ in events 
with $m_T>140$~GeV. The signal is shown for two different masses of the $W'$ boson.
\label{fig:final_plots2}}
\end{figure*}

The contribution from QCD multijet events is estimated using a control sample derived 
from data with the same kinematic cuts. In this sample, the electron candidate fails 
the shower shape requirement. The resulting events are normalized to the data sample. 
The scale factor for the entire QCD multijet sample is adjusted in the low 
reconstructed transverse mass region ($m_T<30$~GeV), which is dominated by QCD multijet 
background events, such that the sum of the \textsc{pythia} Monte Carlo prediction and 
the QCD multijet sample describes the data as shown in Fig. \ref{fig:final_plots2}(a). 
The data are normalized to $W$ boson production and decay in the $e\nu$ mode 
using the $W$ peak region (60~GeV~$<m_T<$~140~GeV, as shown in 
Fig. \ref{fig:final_plots2}(a)) because many efficiency and acceptance errors largely 
cancel in this ratio. We use the theoretical prediction for the $W$ boson production 
cross section $\sigma_W$ $\times$ $B(W\rightarrow e\nu)=2583^{+94}_{-84}$~pb 
from Ref. \cite{hamberg}. 

Jets may be present in conjunction with a $W$ boson due to higher order QCD 
contributions. Since \textsc{pythia} does not properly describe the transverse 
momentum distribution of the $W$ boson in such processes, this spectrum is 
separately reweighted in events with one, two and three jets in order to match 
the distributions observed in the data. This correction affects 10\% of the $W$ 
Monte Carlo events. The sample defined by the selection cuts mentioned above 
contains 452,984 data events compared to 454,000~$\pm$~35,000 events expected 
from SM processes and instrumental backgrounds after applying all corrections.

Two kinds of systematic uncertainties contribute in this analysis.
The uncertainties of the normalization in the $W$ peak region (4\%), 
the cross sections of the SM processes (4-10\%), the electron reconstruction 
efficiency corrections (2\%) and the scale factor for the QCD multijet sample (7\%) 
affect only the global normalization. Uncertainties on the PDFs, electron energy 
scale and resolution, jet energy scale, decay width $\Gamma_W$ of the $W$ boson, 
and the reweighting of the transverse momentum of the $W$ boson lead to changes of 
the shape of the distributions.

In order to study the effect of the electron energy scale and resolution, the electron 
energies have been varied within the known uncertainties. The variations of scale 
and resolution are performed independently. The $\MET$ is recalculated after varying 
the electron energy. The overall uncertainty on the event numbers is large for 
the $W$ sample (4\%), but small for the $W'$ signal ($<$~1\% for 500~GeV~$<m_{W'}<$~1200~GeV). 
The uncertainty of the energy resolution is an order of magnitude smaller than the energy 
scale uncertainty.

In order to study the PDF uncertainty, the Monte Carlo events which have been produced using
CTEQ6L1 PDFs are reweighted to CTEQ6.1M.xx (xx = 0, \ldots, 40), making use of the CTEQ6.1M
PDFs and the 40 error functions \cite{cteq}. The overall uncertainty varies between 3\%
($m_{W'}=500$~GeV) and 8\% ($m_{W'}=1200$~GeV). For the $W$ sample an uncertainty of 3\% is 
derived. The width of the $W$ boson is known to about 2\% \cite{pdg}. This can cause a 
shift ($\sim$~4\%) of the tail of the transverse mass distribution of the $W$ boson.
Finally, the jet energy scale has been varied, and the $\MET$ recalculated. The resulting
uncertainty is below 1\%.

The tail of the spectrum ($m_T>140$~GeV) is now considered to search for $W'\rightarrow e\nu$. 
A good agreement between data and background prediction can be observed as shown in 
Fig. \ref{fig:final_plots2}(b, c). In Table \ref{tab:final_numbers}, the breakdown of the 
individual contributions of the various background processes is given, including expected 
numbers of signal events. Since we do not observe any significant excess in the data, an 
upper limit is set on the production cross section times branching fraction 
$\sigma _{W'} \times B(W'\rightarrow e\nu)$.

\begin{table}
\caption{Event numbers in the data compared to the background prediction
after applying the cut on the transverse mass $m_T>140$~GeV. For the signal and 
background processes statistical and systematic uncertainties are given. 
\label{tab:final_numbers}}
\begin{ruledtabular}
\begin{center}
\begin{tabular}{lrrr}
Process & \multicolumn{1}{c}{Events} & stat & sys\\ \hline
                       \bf Data &  \bf     967 &       &            \\ 
             Sum of backgrounds &          959 &        21 &  90 \\ \hline
            $W\rightarrow e\nu$ &          875 &        20 &  90 \\
       QCD multijet (from data) &           27 &         2 &   2 \\ 
                          other &           57 &         3 &   4 \\ \hline
$W'\rightarrow e\nu$ & & \\
$m_{W'}= 500$~GeV &         1169 &        24 &   86 \\
$m_{W'}= 600$~GeV &          393 &         8 &   32 \\
$m_{W'}= 700$~GeV &          147 &         3 &   13 \\
$m_{W'}= 800$~GeV &           51 &       1.1 &   5.4 \\
$m_{W'}= 900$~GeV &           19 &       0.4 &   2.4 \\ 
$m_{W'}= 1000$~GeV &         7.4 &       0.2 &    1.1 \\
$m_{W'}= 1100$~GeV &         3.4 &       0.1 &    0.5 \\
$m_{W'}= 1200$~GeV &         1.7 &       0.1 &    0.2 \\
\end{tabular}
\end{center}
\end{ruledtabular}
\end{table}

The limit is derived using a binned likelihood for the whole transverse mass 
spectrum with 140~GeV~$<m_T <$~1000~GeV. The individual shape-changing systematic 
uncertainties (up and down variation) enter the limit calculation via individual 
histograms; bin correlations are taken into account.

A Bayesian approach \cite{d0limit} is used to calculate upper limits on the cross 
section for different resonance masses. A Poisson distribution is assumed for the 
number of expected events in each bin of the transverse mass distribution, as well 
as flat prior probabilities for the signal cross sections. The prior for the 
combined signal acceptance and background yields is a multivariate Gaussian with 
uncertainties and correlations described by the corresponding covariance matrix.

\begin{figure}[t]
\def\epsfsize#1#2{0.3#1} 
\centering\epsffile{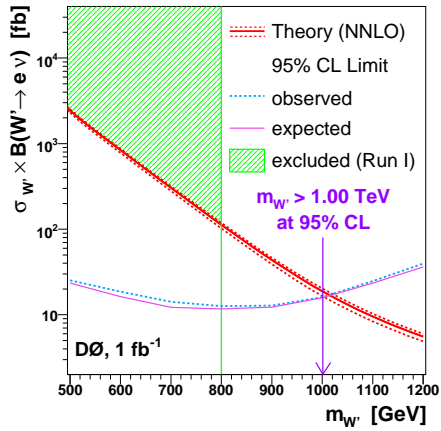}
\caption{The observed and expected 95\% C.L. limits on the cross section as a
function of the mass of the $W'$ boson, including statistical and systematic 
uncertainties. The expected limit assumes a background-only hypothesis.
The theoretical expectation is displayed with its uncertainty.
Also shown is the D0 Run~I limit \cite{oldWp}.
\label{fig:final_limit}}
\end{figure}

The observed and expected 95\% C.L. limits on the production cross section times 
branching fraction $\sigma _{W'} \times B(W'\rightarrow e\nu)$ are shown in 
Fig. \ref{fig:final_limit}. The lower bound of the theoretical cross section 
is used to obtain the mass limit. Hence, an additional heavy charged gauge boson 
with mass below 1.00~TeV is excluded at the 95\% C.L.

In summary, a search for a new heavy charged gauge boson $W'$ decaying to 
an electron and a neutrino has been performed using 1~fb$^{-1}$ of data collected 
with the D0 detector in Run~II. We do not observe an excess in the data, and set 
upper limits on the cross section times branching fraction, which are of the order 
of 10 -- 40~fb for $W'$ boson masses of 500~GeV~$< m_{W'} <$~1200~GeV. 
Further, a lower limit on the mass of the $W'$ boson is derived, assuming that
the new gauge boson as introduced in \cite{Altarelli} has the same couplings 
to fermions as the SM $W$ boson. We exclude a $W'$ boson with $m_{W'}<1.00$~TeV 
at the 95\% C.L. This result represents the most stringent limit on the mass 
of a charged heavy gauge boson beyond the standard model to date.

\input acknowledgement_paragraph_r2.tex

\end{document}

%% file: list_of_authors_r2.tex
%
\author{V.M.~Abazov$^{36}$}
\author{B.~Abbott$^{76}$}
\author{M.~Abolins$^{66}$}
\author{B.S.~Acharya$^{29}$}
\author{M.~Adams$^{52}$}
\author{T.~Adams$^{50}$}
\author{E.~Aguilo$^{6}$}
\author{S.H.~Ahn$^{31}$}
\author{M.~Ahsan$^{60}$}
\author{G.D.~Alexeev$^{36}$}
% ...

\affiliation{\vspace{0.1 in}(The D\O\ Collaboration)\vspace{0.1 in}}
% ...
\affiliation{$^{6}$University of Alberta, Edmonton, Alberta, Canada,
                Simon Fraser University, Burnaby, British Columbia, Canada,
                York University, Toronto, Ontario, Canada, and
                McGill University, Montreal, Quebec, Canada}
% ...
\affiliation{$^{29}$Tata Institute of Fundamental Research, Mumbai, India}
% ...
\affiliation{$^{31}$Korea Detector Laboratory, Korea University, Seoul, Korea}
% ...
\affiliation{$^{36}$Joint Institute for Nuclear Research, Dubna, Russia}
% ...
\affiliation{$^{50}$Florida State University, Tallahassee, Florida 32306, USA}
% ...
\affiliation{$^{52}$University of Illinois at Chicago,
                Chicago, Illinois 60607, USA}
% ...
\affiliation{$^{60}$Kansas State University, Manhattan, Kansas 66506, USA}
% ...
\affiliation{$^{66}$Michigan State University,
                East Lansing, Michigan 48824, USA}
% ...
\affiliation{$^{76}$University of Oklahoma, Norman, Oklahoma 73019, USA}
% ...

%% file: acknowledgement_paragraph_r2.tex
%
We thank the staffs at Fermilab and collaborating institutions, 
and acknowledge support from the 
DOE and NSF (USA);
CEA and CNRS/IN2P3 (France);
FASI, Rosatom and RFBR (Russia);
CAPES, CNPq, FAPERJ, FAPESP and FUNDUNESP (Brazil);
DAE and DST (India);
Colciencias (Colombia);
CONACyT (Mexico);
KRF and KOSEF (Korea);
CONICET and UBACyT (Argentina);
FOM (The Netherlands);
Science and Technology Facilities Council (United Kingdom);
MSMT and GACR (Czech Republic);
CRC Program, CFI, NSERC and WestGrid Project (Canada);
BMBF and DFG (Germany);
SFI (Ireland);
The Swedish Research Council (Sweden);
CAS and CNSF (China);
Alexander von Humboldt Foundation;
and the Marie Curie Program.
%

%% file: prl.bbl
\begin{thebibliography}{99}
\bibitem{phi}
The azimuth angle $\phi$ is defined in the plane perpendicular to the beams, with $\phi=\pi/2$ pointing up.
\end{thebibliography}
